\documentclass[aps,twocolumn,amsmath,prl,showkeys]{revtex4-1}
\usepackage{graphicx,color}
\usepackage{subfigure}
\usepackage{amsmath}
\usepackage{bibentry}

\begin{document}

\title{Universal Extraction of Quantum Critical Exponents and Phase Transitions
via Tailored Hilbert Space}

\author{Ye Xiong} 
\email{xiongye@njnu.edu.cn}
\affiliation{Institute of Theoretical Physics, Nanjing Normal University, Nanjing 210023,
P. R. China}

\begin{abstract}

   Finite-size scaling and the renormalization group form the central toolkit
   for analyzing quantum phase transitions (QPTs). In this Letter, we introduce
   a novel Hilbert-space tailoring scheme to probe quantum critical phenomena.
   Applied to the second-order QPT of the one-dimensional (1D) XY model, our method
   yields precise critical points and exponents on lattices containing merely 50
   unit cells. We further establish the universal applicability of this
   framework via investigations of the Berezinskii-Kosterlitz-Thouless
   transition in the 1D XXZ chain: critical parameters are recovered
   with as few as 12 lattice sites. This technique may open an alternative,
   efficient route to universally characterize QPT across many-body lattice systems.

\end{abstract}

\maketitle

\section{Introductions.}

Quantum mechanics is built upon two foundational constructs: Hilbert space and
the operators that act within it. Nevertheless, nearly all established
techniques for investigating quantum phase transitions (QPTs) center on the
operator side of the formalism \cite{sachdev2011qpt,
chaikin2000condensed, abrikosov1963manybody,
vojta2003qptreview, schollwoeck2011dmrg} . Green’s functions—the prevailing perturbative
toolset—illustrate this paradigm, capturing the linear response of quantum
many-body systems to local perturbations applied at spacetime coordinates
$(r',t')$. Nearly every core result shaping our knowledge of QPTs, spanning
correlation lengths and order parameters alike, originates from this
operator-based formalism\cite{abrikosov1963manybody, vojta2003qptreview,
kadanoff1963green, CARR, PFEUTY197079,PhysRevB.107.045124}.

On the contrary, a far narrower class of theoretical approaches directly
manipulates Hilbert space itself. Spatial bipartitioning serves as a
representative illustration: this procedure artificially partitions the full
Hilbert space into two disjoint subsystems, and its implementation led to the
discovery of the entanglement entropy area law. For one-dimensional (1D) systems
tuned to criticality, the entanglement entropy that quantifies quantum
cross-subsystem correlations exhibits logarithmic scaling with total system
size \cite{RevModPhys.82.277,Calabrese2004, Lieb1972, PhysRevLett.96.010404,
PhysRevLett.96.100503, PhysRevLett.105.050502,
PhysRevLett.96.110404,Pastawski2015, David2015}.

Both families of approaches are hampered by finite-size artifacts. Spectral
smearing and energy-level shifts \cite{PhysRevLett.28.1516} inherent to finite lattices demand exhaustive
finite-size scaling calculations on very large systems to reliably pinpoint
critical points and extract accurate critical exponents. The steep computational
cost of such large-scale computations imposes severe practical limitations,
sustaining long-standing open disputes over the categorization of QPT behavior
in many canonical lattice models\cite{RevModPhys.66.763, RevModPhys.68.13,
RevModPhys.78.17}.

Inspired by the bipartition that to manipulating the Hilbert space, we develop a new
method, which is called tailoring the Hilbert space (THS), to study QPT. The most
advantage of the method is one can obtain the critical point and the critical
exponents in small system without the finite size scaling.

The model Hilbert space is tailored by introducing an additional projection term in the Hamiltonian
\begin{equation}
H_W = H + W_1 |\Psi_1\rangle\langle\Psi_1|,
\end{equation}
where $H$ denotes the original Hamiltonian for the QPT system under
investigation. The parameters $W_1$ are formally infinite but can be numerically
approximated by a large energy scale, which is several orders of magnitude
larger than all characteristic energy scales of $H$. Here, $\{|\Psi_1\rangle\}$
defines the state set implementing the Hilbert-space tailoring scheme.
Obviously, $|\Psi_1\rangle$ is an eigenstate of $H_W$ with
divergent eigenvalue $W_1$. In this construction, finite-energy bands remain
governed by the original Hamiltonian $H$, while confined within the truncated
Hilbert subspace $(I-|\Psi_1\rangle\langle\Psi_1|)$ that excludes the
subspaces spanned by $\{|\Psi_1\rangle\}$.

The conventional one-dimensional (1D) spatial bipartition scheme can be
naturally incorporated within this framework by choosing the state
$\Psi_1(x)=\delta(x-x_0)$, with $x_0$ labels the central lattice site of the
system. For systems with only nearest-neighbor hopping, the tailored Hilbert
space effectively decouples the original chain into two disconnected subchains.
We further demonstrate that generalized choices of $\{|\Psi_1\rangle\}$ provide
versatile tools for probing quantum critical behaviors encoded in $H$.

Prior implementations of Hilbert-space projection generally rely on
operator-based constraints. A canonical example is the strongly interacting
Hubbard model with large on-site repulsion $U$, where the low-energy effective
Hilbert space is constructed by projecting out all doubly occupied
subspaces\cite{ANDERSON1973153,Chao_1977,PhysRev.147.392,PhysRev.149.491}.
This conventional projection scheme fundamentally differs from our THS
approach: standard operator-based projections eliminate extensive
state manifolds modified by interaction constraints, whereas our tailoring
method only excludes a minimal set of deliberately designed state
$|\Psi_1\rangle$.

In our previous work, we applied THS to the one-dimensional
Anderson disorder model using a uniform extended state $\Psi_1(x)=1$\cite{Xiong_2023}. The
tailored framework modifies the original localized eigenstates, yielding
partially extended wave function. While these states retain the short-distance
exponential decay characteristic of conventional Anderson localization, their
long-range tails become fully extended due to hybridization with the tailored
state $|\Psi_1\rangle$. This phenomenon motivates the present study of QPTs. 
Since ground states belonging to distinct
quantum phases occupy different Hilbert-space regions and exhibit distinct
overlaps with the tailored state $|\Psi_1\rangle$, THS should
naturally amplifies phase distinctions and renders quantum critical behavior
more identifiable.

\section{Second-Order Quantum Phase Transitions in the One-Dimensional XY Model
within the Tailored Hilbert-Space Framework}

We study the spinless fermion Hamiltonian
\begin{equation}
\begin{aligned}
H = \sum_i \Big[ & \left(c^\dagger_{i+1} c_i - d^\dagger_{i+1} d_i + \text{h.c.}\right)
+ \gamma \left(c^\dagger_{i+1} d_i - d^\dagger_{i+1} c_i + \text{h.c.}\right) \\
&+ \frac{h}{2} \left(c^\dagger_i c_i - d^\dagger_i d_i\right) \Big],
\end{aligned}
\end{equation}
where $\text{h.c.}$ denotes the Hermitian conjugate term. By imposing
particle-hole symmetry in the Nambu representation through the mapping $d_i =
c^\dagger_i$, this model is exactly mapped to the one-dimensional quantum XY
model via the Jordan--Wigner transformation. Importantly, we do not impose
particle-hole symmetry a priori in the bare Hamiltonian, since THS
state $|\Psi_1\rangle$ can either preserve or explicitly break
this symmetry. The system hosts a Ising second-order quantum phase transition (QPT)
located at the critical point $h=1$.

\begin{figure}[ht]
\includegraphics[width=0.45\textwidth]{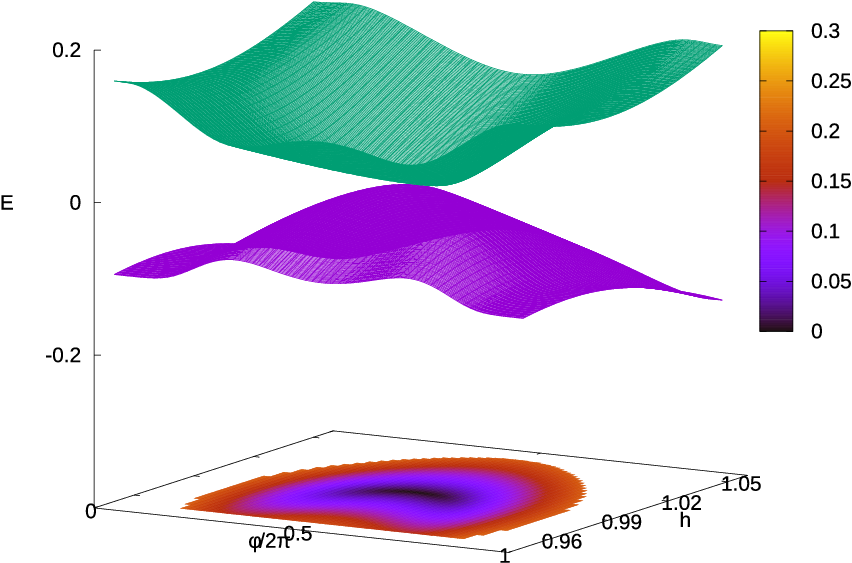}
\caption{\label{fig1} Eigenenergies of the tailored Hamiltonian $H_W$ as
functions of $h$ and $\phi$. The bottom palette illustrates the energy gap
between the two states. 
Calculations are performed at $\gamma=0.5$ and fixed tailoring strength $W_1=10^5$ throughout this work. 
Only low-energy states near zero energy are displayed. A characteristic cone-shaped dispersion emerges near 
the critical point $h=1$, clearly signaling the quantum critical behavior.}
\end{figure}

In Fig. \ref{fig1}, we show the energy spectrum of $H_W$ against $h$ and $\phi$ for an $N=25$ ring, with $\phi$ labeling the threaded magnetic flux. The THS is built from a random tailored state
\begin{equation}
|\Phi_1\rangle = \sum_i f_i c^\dagger_i |\text{Vac}\rangle,
\end{equation}
where $f_i \in [-0.5,0.5]$ follow uniform random distributions, and
$|\text{Vac}\rangle$ stands for the vacuum state. A distinct conical band
crossing arises at $(h=0.998,\phi=\pi)$, enabling direct extraction of the
critical point. Finite-size spectral rounding shifts the apparent critical
position at order $1/N$; however, this minor deviation is much weaker than the
pronounced susceptibility divergence seen in conventional finite-size scaling,
which validates the reliability of our THS framework.

Generic condensed-matter systems display avoided level crossings without
protective symmetries. Here, $|\Psi_1\rangle$ breaks the particle-hole symmetry
of bare $H$, leaving no conventional symmetry to sustain the cone. Still, the
conical feature proves highly robust across various setups: different random
$|\Psi_1\rangle$ samples, two-state THS constructions restoring particle-hole
symmetry, and translational-symmetry-breaking hopping perturbations that erase
all cone signatures in the bare finite-size spectrum. No standard mechanism
accounts for this robust spectral structure. In the next section, we prove that
THS engineering generates an emergent effective symmetry shielding the conical
dispersion.

\begin{figure}[ht]
\includegraphics[width=0.45\textwidth]{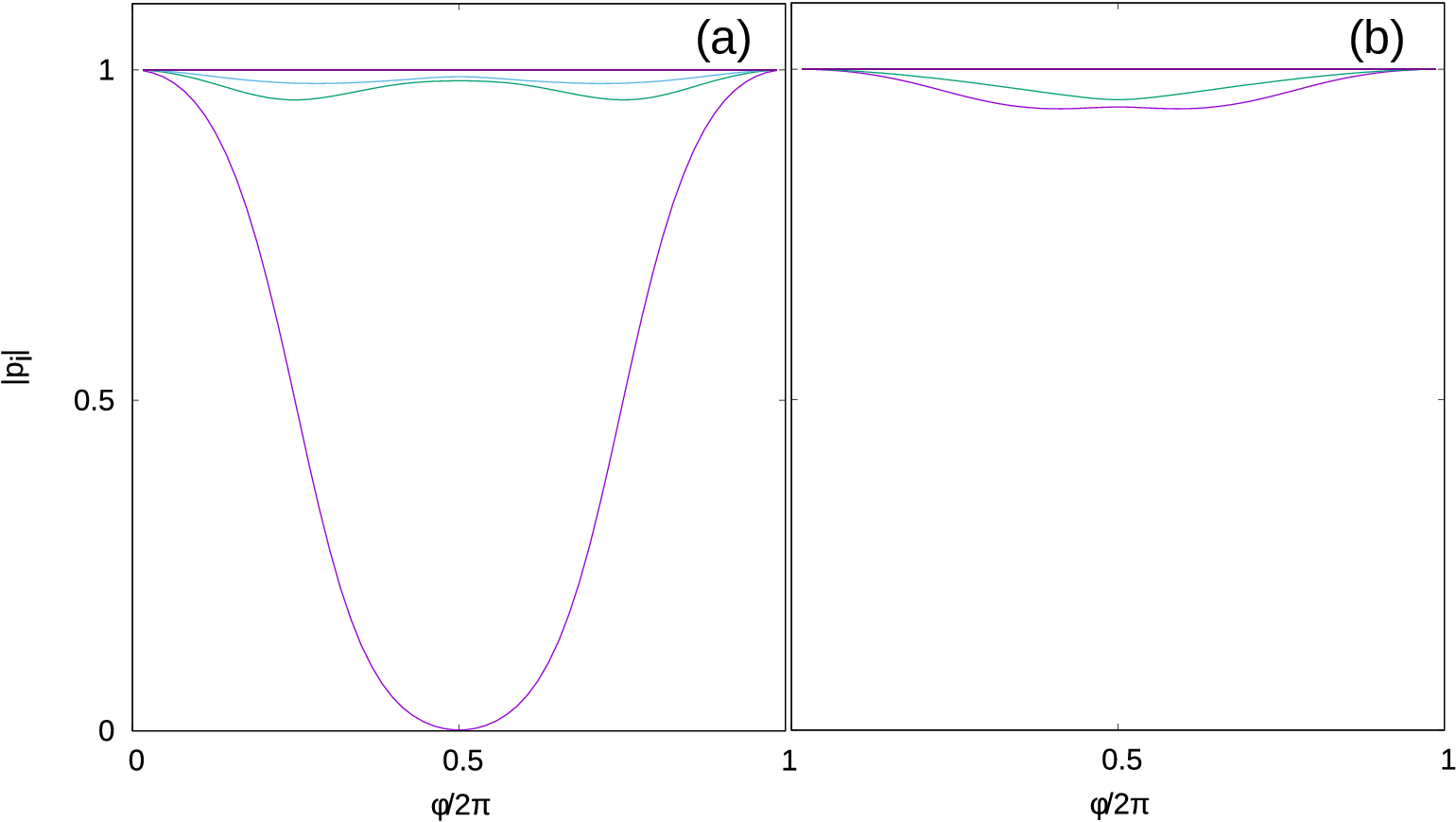}
\caption{\label{fig2} Eigenvalues of the projection operator that maps low-energy eigenstates onto the subspace spanned by the $\phi=0$ lower-band states. Panels (a) and (b) correspond to $h=0.6$ and $h=1.4$, respectively.}
\end{figure}

In Fig. \ref{fig2}, we present the eigenvalues of the projection operator
$\sum_{E_i<0} |\alpha_i(\phi)\rangle \langle \alpha_i(\phi)|$ projected onto the
subspace spanned by $\{|\alpha_i(\phi=0)\rangle\}$ with $E_i<0$.  Here, $E_i$
denotes the eigenenergy of $H$ within the THS, and $|\alpha_i\rangle$ the
corresponding eigenstate.  These curves illustrate the evolution of the
negative-energy subspace ($E<0$) as a function of flux $\phi$ deep inside two
distinct phases at $h=0.6$ and $h=1.4$. The chain size is same as that in Fig.
\ref{fig1}.

In Fig. \ref{fig2}(a), one eigenvalue smoothly drops to zero and recovers at
larger $\phi$, whereas all eigenvalues remain finite throughout Fig.
\ref{fig2}(b).  At $h=0.6$, even though all low-band energies stay negative, one
state continuously mixes into the upper-band subspace as $\phi$ varies.
Correspondingly, one upper-band state simultaneously migrates into the
lower-band subspace (not plotted).  The paired exchange of these two states,
combined with the fixed tailored state $|\Psi_1\rangle$, mimics rotation about a
fixed axis in three dimensions.  The winding number characterizing this
rotational behavior carries topological origin, anchored by the invariance of
$|\Psi_1\rangle$.  It can only lose its well-defined nature when the two
migrating states become indistinguishable in the two bands, i.e., when the band
gap between upper and lower bands vanishes.

This topological invariant differs fundamentally from conventional winding
numbers in established topological physics in three key aspects.  First, it is
defined over the full Hilbert space, independent of real-space or momentum-space
representations.  Second, its stability is guaranteed not by intrinsic
symmetries of the bare Hamiltonian $H$, but by an emergent symmetry unique to
THS construction, whose existence relies on the fixed tailored state
$|\Psi_1\rangle$.  Third, conventional topological invariants become ill-defined
precisely at quantum critical points. In contrast, our winding number loses its
validity when the two bands become indistinguishable. Band-gap closing
accompanies both phenomena.

We now address a natural follow-up question: can the THS framework offer a novel
pathway for extracting critical exponents? The answer is affirmative. We begin
with the correlation length $\xi$, defined via the spin-spin correlation
function
\begin{equation}
\langle G| \hat m_{x,i} \hat m_{x,j} |G\rangle - \langle G| \hat m_{x,i} |G \rangle\langle G| \hat m_{x,j} |G \rangle
\propto e^{-|i-j|/\xi},
\end{equation}
where the local order parameter at site $i$ reads $\hat m_{x,i}=\hat{K}
c^\dagger_i + d^\dagger_i \hat{K}$, $\hat K$ denotes the Jordan–Wigner string
operator, and the ground state takes the form $|G\rangle
=\frac{1}{\sqrt{2}}\big(|{\rm even}\rangle + |{\rm odd}\rangle\big)$, a coherent
superposition of the even- and odd-particle-number ground manifolds. Standard
analyses of correlation lengths commonly neglect the Jordan–Wigner string $\hat
K$. To further simplify calculations, we instead examine correlations of the
local observable $\hat m_{z,i}= c^\dagger_i c_i -d^\dagger_i d_i$.  The
expectation value $m_{z,i} =\langle GG| \hat m_{z,i} |GG \rangle$ admits
straightforward evaluation on the many-body ground state $|GG\rangle =
\prod_{E<0} \alpha^\dagger_{E} |\text{Vac}\rangle$, a Slater determinant constructed from
all single-particle eigenstates with negative energy. 
By Wick’s theorem, the two-point correlation of $\hat
m_z$ decays as $e^{-2|i-j|/\xi}$.

\begin{figure}[ht]
\includegraphics[width=0.5\textwidth]{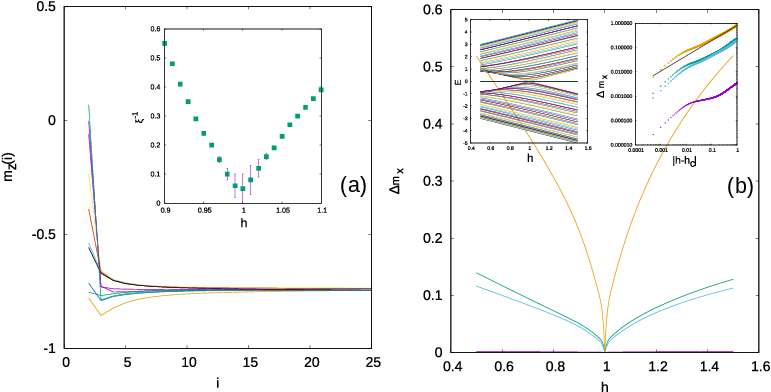}
\caption{\label{fig3} (a) Site-resolved $m_z$ calculated in the THS for an
ensemble of $|\Psi_1\rangle$ on an $N=50$ ring. All curves saturate to a
constant via exponential decay. The inset displays correlation-length scaling
$\xi \propto |h-h_c|^{-\nu}$ with $\nu=1$.  (b) Left inset: THS energy spectrum
hosting a zero-energy in-gap state. Main panel: Curves from bottom to top show
the zero-energy contribution $\Delta m_x$ at $N=100,200,400,1000$, whose scaling
with system size $N$ verifies the extended nature of this midgap state. Right
inset: Double-logarithmic plot of $\Delta m_x$ against $h-h_c$, demonstrating
the power-law behavior $\Delta m_x \propto (h-h_c)^{0.625}$. The solid line acts
as a guide to the eye for the measured exponent.}
\end{figure}

Correlations of $\hat m_{z,j}$ can be readily accessed when the tailored state
$|\Psi_1\rangle$ is localized near site $i$. Physically, $\hat m_{z,j}$
quantifies how the local Hilbert-space deformation induced at site $i$
propagates across the lattice to site $j$, which exactly encodes the correlation
behavior of the system. In Fig. \ref{fig3}(a), we plot $m_{z,i}$
versus lattice index $i$ for multiple $|\Psi_1\rangle$s computed within the THS.
The tailored state is set to $|\Psi_1\rangle= (f_1 c^\dagger_1 + g_1
d^\dagger_1) |\text{Vac}\rangle$, with $f_1,g_1$ sampled uniformly from
$[-0.5,0.5]$.`

The curves demonstrate that $m_{z,i}$ saturates at large distances, while its
local distortion decays exponentially with lattice index $i$, enabling
extraction of the correlation length $\xi$.  The inset in Fig. \ref{fig3}(a) presents the
inverse correlation length $\xi^{-1}$ plotted against $h$.  Fitting to the
critical scaling relation $\xi \propto |h-h_c|^{-\nu}$ yields a correlation-length
exponent $\nu=1$.  Finite-size rounding blurs the precise value of $\xi$ near
the critical point $h_c$ owing to the small ring geometry; this artifact
diminishes as system size grows.  Notably, our THS scheme reliably recovers the
critical exponent even for parameters far from $h_c=1$, in stark contrast to
conventional finite-size computations.  This advantage becomes even more
prominent when evaluating subsequent critical exponents, whose underlying
mechanism we elaborate on below.

We now turn to the computation of the critical exponent $\beta$, which governs
the order-parameter scaling $m_x \propto |h-h_c|^{\beta}$. Rather than
evaluating $m_x$ directly, we extract the shift $\Delta m_x$ induced by THS
modification.  We adopt a tailored state of the form
$$
|\Psi_1\rangle = \sum_i f_i (c^\dagger_i+d^\dagger_i)|\text{Vac}\rangle,
$$
where each $f_i$ is a uniform random variable sampled over $[-0.5,0.5]$. The THS
energy spectrum is plotted in the left inset of Fig. \ref{fig3}(b). A
distinctive feature introduced by the tailored subspace is a persistent
zero-energy state present for all values of $h$. This state differs
fundamentally from Majorana edge modes, which only emerge within the topological
phase. More remarkably, this in-gap state is fully extended across the ring, a
direct consequence of the THS construction.

The modified ground state takes the form $|G\rangle \to
(\alpha^\dagger_{E=0}+1)|G\rangle$, a coherent superposition mixing the
zero-energy mode with the original low-energy ground manifold (itself a
superposition of even- and odd-particle-number states). We find the resulting
ground-state energy shift obeys $\Delta E \propto |h-h_c|$, consistent with the
dynamical critical exponent $z=1$.  The order-parameter variation
$$ \Delta m_x = \langle \text{Vac}| \sum_i \hat m^x_i \alpha^\dagger_{E=0}
|\text{Vac}\rangle $$
admits straightforward analytical evaluation. We omit the
usual $1/N$ normalization of $\Delta m_x$ to explicitly demonstrate the extended
character of the zero-energy state in our plots.

Figure \ref{fig3}(b) presents $\Delta m_x$ as a function of $h$, recovering the
power-law scaling $\Delta m_x \propto |h-h_c|^{0.625}$. We stress that this
clean scaling behavior is already observable on relatively small rings ($N=100$)
and at parameters far from $h_c$. As the system size increases, the scaling
regime broadens toward the critical point and the power-law signature becomes
more pronounced, a trend also visible in Fig. \ref{fig3}(a). We further
demonstrate that this favorable feature persists near the
Berezinskii-Kosterlitz–Thouless (BKT)
transition of the XXZ model, and we elaborate on the underlying physical
mechanism in the subsequent section.

We extend the generalized Landau theory for second-order phase transitions to
relate the measured exponent $0.625$ to the standard critical exponent
$\beta=1/8$.  We start from the Landau free energy $\tilde E$, minimized with
respect to the order parameter $m_x$:
\begin{equation}
\tilde E = a\, \text{sign}(h-h_c) \left|h-h_c\right|^{1/4} (m_x)^2 + b(m_x)^4,
\end{equation}
where $a,b>0$ are positive constants, and $\text{sign}(x)$ yields $+1$ for
$h>h_c$ and $-1$ for $h<h_c$. The power $|h-h_c|^{1/4}$ originates from the
known exponent $\beta=1/8$.

The THS construction introduces an energy shift $\Delta \tilde E$, and the modified total energy remains stationary under variations of the shifted order parameter $m_x+\Delta m_x$. This yields the cross term
$$
\Delta \tilde E \propto \left|h-h_c\right|^{1/4} m_x \Delta m_x.
$$
Substituting our earlier result $\Delta \tilde E \propto |h-h_c|$ into this relation directly gives $\Delta m_x \propto |h-h_c|^{5/8}$, matching the measured exponent $0.625$ shown in Fig. \ref{fig3}(b).

We have uncovered a novel feature unique to the THS framework: critical exponents can
be reliably extracted at parameter values far from the quantum critical point
$h_c$. This favorable property stands in stark contrast to conventional
finite-size scaling approaches for QPTs. Standard
methods demand large system sizes to accurately locate the critical point, and
precise exponent extraction is only feasible within a narrow window very close
to $h_c$. Even tiny uncertainties in the inferred critical-point position
significantly degrade the resulting exponent estimates. In contrast, our THS
scheme yields robust critical exponents well away from $h_c$. Consequently,
exponent calculations remain largely unaffected even when the critical point
itself is poorly resolved on small lattices.

\section{BKT Transition of the XXZ Model within the THS Framework}

\begin{figure}[ht]
\includegraphics[width=0.45\textwidth]{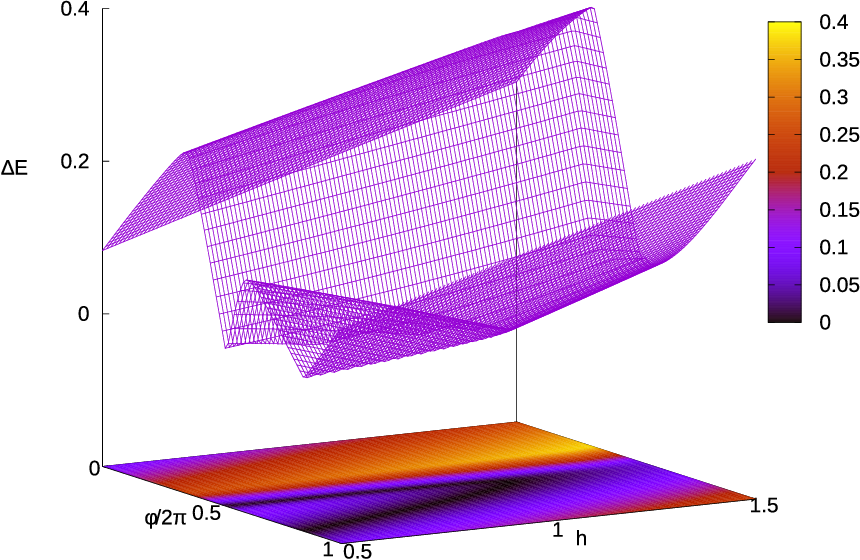}
\caption{\label{fig4} The gap between the ground state and first excited state
energies in THS.}
\end{figure}

We have demonstrated the utility of the THS scheme for second-order quantum
phase transitions in the XY model. A natural question arises: can this framework
be generalized to other classes of QPTs? We address this by applying our
approach to the XXZ model hosting a BKT transition.

We consider a one-dimensional fermionic model defined as
\begin{equation}
H=\sum_i c^\dagger_i c_{i+1} +\text{h.c.} + \Delta \sum_i \left(n_i-\tfrac12\right)\left(n_{i+1}-\tfrac12\right),
\end{equation}
where $n_i = c^\dagger_i c_i$ denotes the fermion number operator at site $i$.
This fermionic Hamiltonian can be mapped onto the spin XXZ model via the inverse
Jordan–Wigner transformation. The model exhibits a first-order quantum phase
transition at $\Delta=-1$ and a BKT transition at $\Delta=1$.  For $\Delta>1$,
the system resides in an antiferromagnetic phase featuring spontaneous
$Z_2$ symmetry breaking. The region $-1<\Delta<1$ corresponds to a
gapless phase lacking long-range magnetic order\cite{sachdev2011qpt,
Baxter1973,JPSJ.62.3774}.  Within the THS framework, the
first-order transition remains intact; we therefore concentrate our subsequent
analysis on the BKT transition.

We analyze the THS energy spectrum as a function of interaction strength
$\Delta$ and threading magnetic flux $\phi$.  Simulations are performed on a
small ring of length $N=12$, which permits full exact diagonalization. We find
that only tailored states of the form $$ |\Psi_1 \rangle = \sum_i f_i e^{I\pi
i/2} c^\dagger_i |\text{Vac}\rangle $$ yield reliable probes of the system,
where each $f_i$ is a random real coefficient and $I$ denotes the imaginary
unit. We stress that the introduced phase factor generates a 4-fold periodicity,
doubling the intrinsic period-2 response of the antiferromagnetic phase.

Figure \ref{fig4} displays the energy gap separating the ground state and the
first excited state for uniform weights $f_i=1$. For $\Delta_c < 1.2$, the gap
closes twice as $\phi$ varies. When $f_i$ are randomized, the overall spectral
profile remains qualitatively unchanged, while the apparent critical coupling
$\Delta_c$ fluctuates around 1. This deviation is expected given the very small
system size, which prevents precise localization of the true critical point. We
now investigate whether critical exponents can still be reliably extracted under
such finite-size limitations.

\begin{figure}[ht]
\includegraphics[width=0.45\textwidth]{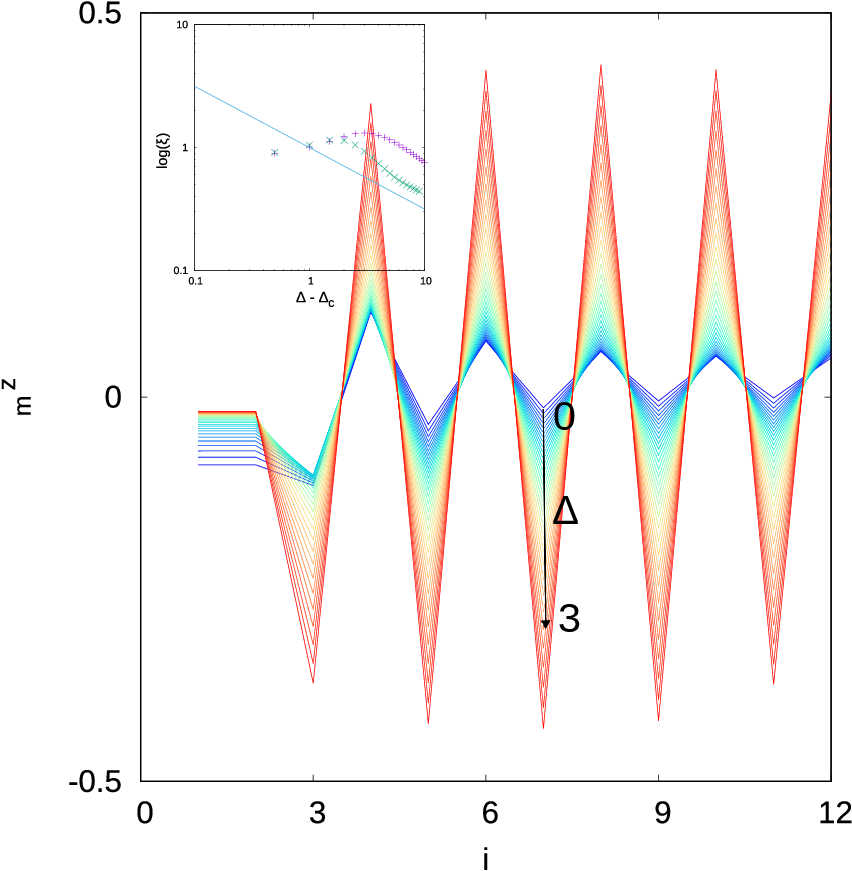}
\caption{\label{fig5} Site-resolved $m_z$ versus lattice index $i$ for an
$N=12$ open-boundary chain. $\Delta$ is taken from $0$ to $3$. 
The tailored state $|\Psi_1\rangle$ is localized on the
leftmost sites $i=1$ and $i=2$. The inset, arranged from top to bottom, presents
the logarithmic correlation length for system sizes $N=12$ and $N=16$,
respectively. The solid line serves as a guide to the eye for the scaling
behavior $x^{-0.5}$.}
\end{figure}

Figure \ref{fig5} plots the expectation value of $\hat{m}^z_i=2c^\dagger_i c_i
-1$ evaluated with the tailored state $|\Psi_1 \rangle = \big(c^\dagger_1 +
c^\dagger_2\big)|\text{Vac}\rangle$. We adopt open boundary conditions to
visualize how correlations propagate across the small chain after local
perturbation by the THS at the left edge.

Near $\Delta=0$, $m^z_i$ exhibits mild oscillations yet tends to decay and saturate to
zero at the right boundary. For $\Delta>1$ — where precise extraction of the
transition point is infeasible given the tiny system size — $m^z_i$ saturates to
values characteristic of the antiferromagnetic phase at the right.

Crucially, the tailored state $|\Psi_1\rangle$ itself does not explicitly break
the $Z_2$ symmetry. This observation leads to a central conclusion:
spontaneous symmetry breaking associated with the quantum phase transition can
emerge within the THS framework even on finite lattices. Conventional quantum
phase transition theory asserts that symmetry breaking only occurs strictly in
the thermodynamic limit of infinite system size. For finite-size calculations,
spontaneous symmetry breaking is absent, so an explicit infinitesimal
symmetry-breaking field must be introduced artificially. Such an external bias
contaminates weak quantum fluctuations far from the critical point and
eliminates the ability to reliably extract critical exponents in that regime.

By contrast, our two model systems demonstrate this unique advantage of THS.
While the first XY model does not directly yield the bare order parameter, it
accesses the order-parameter shift $\Delta m_x$; the XXZ model further
corroborates this behavior. In both cases, spontaneous symmetry breaking arises
on finite chains purely via Hilbert-space tailoring, with no modification to the
original Hamiltonian. As a direct consequence, critical exponents remain
accessible over a broad parameter range far from the quantum critical point.

We now extract the correlation length $\xi$ from the response of the far-end
order parameter $\delta m^z_N$ induced by a tiny perturbation to the tailored
state. Specifically, we deform the reference state $|\Psi_1\rangle =
\big(c^\dagger_1 + c^\dagger_2\big)|\text{Vac}\rangle$ to $1.001\,c^\dagger_1 +
c^\dagger_2|\text{Vac}\rangle$, with the correlation length defined via
$$
\xi \propto -\frac{1}{\log(\delta m^z_N)}.
$$

The inset of Fig. \ref{fig5} presents a double-logarithmic plot of
$\log(\xi/\xi_{\Delta\to\infty})$ against $\Delta-\Delta_c$, where the reference
length $\xi_{\Delta\to\infty}$ is numerically evaluated at $\Delta=600$ and the
critical coupling is set to $\Delta_c=1$. Data for system sizes $N=12$ and
$N=16$ are both displayed, and the solid line serves as a guide to the eye for
the power-law scaling $(\Delta-\Delta_c)^{-0.5}$.

We clearly recover the characteristic BKT form $\xi \propto
\exp\big(k/\sqrt{\Delta-\Delta_c}\big)$ over the regime $\Delta-\Delta_c>5$.
This valid scaling window broadens and extends closer to $\Delta_c$ when the
chain length is increased to $N=16$. Since the usable scaling region lies far
from the critical point, consistent scaling behavior remains observable even if
the apparent critical coupling shifts to $\Delta_c=1.2$.

\section{Conclusions and Outlook}

We establish THS as a novel framework for
investigating quantum critical phenomena. We have successfully applied this
approach to both the XY and XXZ models, directly extracting critical points and
critical exponents using only minimal finite lattices—without relying on
conventional finite-size scaling techniques. A key advantage is that spontaneous
symmetry breaking naturally emerges within finite systems under the THS
construction, eliminating the need for artificial infinitesimal
symmetry-breaking fields. Consequently, signatures of the quantum phase
transition (QPT) persist over parameter regimes far from the critical point,
rendering our scheme highly promising for probing poorly understood QPTs whose
underlying mechanisms remain under active debate.

The present Letter restricts analysis to energy spectra and order-parameter
responses. Since spatially localized perturbation constitutes a special subclass
of THS constructions, a natural future direction is to characterize system-size
dependence of entanglement entropy within general THS setups. The tailored
reference state $|\Psi_1\rangle$ introduces an extra tunable degree of freedom
to probe such entanglement observables, which promises to deepen our
understanding of area laws, conformal field theory, and quantum criticality.

This work exclusively addresses zero-temperature QPTs, yet the THS paradigm can
be generalized to finite-temperature thermodynamic phase transitions by
tailoring the Hilbert space of thermal density matrices. Viewed from this
perspective, THS offers a potentially universal formalism applicable to all
classes of phase transitions.

Our analysis of the XY model further reveals that THS engenders a novel symmetry
class together with an accompanying topological invariant defined intrinsically
on the modified Hilbert space. This invariant furnishes a unified description
for both clean crystalline lattices and disordered systems, and the emergent
symmetry enlarges the known classification table of topological
materials—converting phases classified as trivial under standard schemes into
topologically nontrivial ones. A calculations on the one-dimensional
SSH model reproduce phenomenology analogous to that displayed in Figs.
\ref{fig1} and \ref{fig2}, demonstrating that THS provides a unified platform to
characterize both conventional second-order QPTs and topological QPTs.

For the XXZ model, our numerical tests are limited to chains of length $N=16$.
Combining THS with advanced many-body numerical algorithms would yield
higher-precision determinations of the critical coupling $\Delta_c$, a
worthwhile extension for subsequent work.

The Hilbert space forms the fundamental foundation of quantum mechanics;
controlled tailoring of this space unlocks a rich spectrum of unforeseen
phenomena across all branches of condensed-matter physics. One striking example
is the extended in-gap state identified in Fig. \ref{fig3}, a feature entirely
absent in standard untreated Hamiltonians without THS modification.

At present, constructing most tailored states $|\Psi_1\rangle$ requires
engineered long-range hopping terms, which poses a substantial challenge for
experimental implementation. Among existing experimental platforms, ultracold
atomic systems stand as the most promising candidate. Certain configurations,
such as the ring geometry illustrated in Fig. \ref{fig2}, present additional
complications: threading magnetic flux $\phi$ around the ring induces Peierls
phases on all long-range hopping bonds, which inherently modifies the tailored
state $|\Psi_1\rangle$ as $\phi$ is tuned. Nevertheless, this experimental
limitation does not invalidate our theoretical treatment, where we adopt a fixed
$|\Psi_1\rangle$ for controlled analysis.

{Acknowledgments.---} 
The work was supported by the 
National Foundation of Natural Science in China Grant Nos. 10704040.

\bibliographystyle{apsrev4-1}
\bibliography{main}

\end{document}